\documentclass[a4,oneside,12pt]{article}
\usepackage{array,longtable}\setlongtables
\usepackage{amsfonts}
\usepackage{amsmath}
\usepackage{amsthm}
\usepackage{verbatim}

{\catcode `\@=11 \global\let\AddToReset=\@addtoreset}
\AddToReset{equation}{section}

\newcommand{\MM}{\scriptscriptstyle}  

\def\3norm{|\!|\!|}

\newtheorem{lemma}{Lemma}[section]
\newtheorem{theorem}[lemma]{Theorem}

\newtheorem{corollary}[lemma]{Corollary}
\theoremstyle{definition}

\theoremstyle{definition}
\newtheorem{remark}[lemma]{Remark}

\newcommand{\N}{\mathbb{N}}

\newcommand{\R}{\mathbb{R}}

\newcommand{\newpar}{\par}\parindent =0pt\parskip=3pt\textheight = 615pt
\renewcommand{\L}{{L}}
\newcommand{\Id}[1]{{\rm I\kern-2pt I_{#1}}}
\renewcommand{\hbar}{{\displaystyle\bar{\phantom{x}}\kern-6pt h}}

\begin{document}

\thispagestyle{empty}
\begin{center}
\fontsize{20}{22}
\selectfont
\textbf{On global classical solutions of the time-dependent von Neumann equation
       for Hartree-Fock systems}
\end{center}
\vskip 1 cm
\begin{center}
\fontsize{12}{12}
\selectfont
A. Arnold${}^{a)}$, R. Bosi${}^{a)}$,
S. Jeschke${}^{b)}$, and E. Zorn${}^{b)}$
\end{center}

 ${}^{a)}$ Institut f\"ur Numerische Mathematik, Universit\"at M\"unster, 
 Einsteinstr.~62, D-48149 M\"unster, Germany,\\
e-mail: anton.arnold@math.uni-muenster.de, bosi@math.uni-muenster.de,
\\
 ${}^{b)}$ Institut f\"ur Mathematik, Technischen Universit\"at Berlin, Stra\ss e des 17.\ Juni 136, 
D-10623 Berlin, Germany,\\
e-mail: sabina@math.tu-berlin.de, erhard@math.tu-berlin.de.

\begin{center}
version: 12 May 2003
\end{center}
\vskip 1 cm
\begin{center}
\textbf{Abstract}
\end{center}

\emph{This paper is concerned with the well-posedness analysis of the Hartree-Fock system modeling
the time evolution of a quantum system comprised of \linebreak 
fermions. We consider quantum states with
finite mass and finite kinetic energy, and the self-consistent potential is the unbounded Coulomb
interaction.  This model is first
formulated as a semi-linear evolution problem for the one-particle density matrix operator lying
in the space of Hermitian trace class operators. Using semigroup techniques and generalized
Lieb-Thierring inequalities we then prove global existence and uniqueness of mild and classical
solutions. To this end we prove that the quadratic Hartree-Fock terms are locally Lipschitz in the
space of trace class operators with finite kinetic energy.\\
Technically, the main challenge stems from considering the model as an evolution problem for
operators. Hence, many standard tools of PDE-analysis (density results, e.g.) are not readily
available for the density matrix formalism.
}

\newpar


\newpage

\renewcommand{\thesection}{\Roman{section}}
\section{Introduction}\label{sec:introduction}
\renewcommand{\thesection}{\arabic{section}}

The time dependent Hartree-Fock theory provides approximate evolution models for many-body quantum
systems comprised of fermions, as it accounts for the Pauli exclusion principle. It was first derived by
Dirac \cite{Di30} and simplified by Slater \cite{Sla50}. The {\it Hartree-Fock
system} is a non-linear evolution equation for the one-particle density matrix operator
$\hat{\varrho}(t)$. It has the form of a {\it von Neumann equation}:
\begin{equation}\label{op_level_initial_value_problem}
\begin{array}{lll}
i \; \hat{\varrho}_{t} (t)
                        & = & [\hat{H}(t), \hat{\varrho} (t)] \;, \quad t > 0 \\ [1ex]
\hat{\varrho} (0)       & = & \hat{\varrho}^{\MM I}
\end{array}
\end{equation}
for a given initial value $\hat{\varrho}^{\MM I}$, and $[.,.]$ denotes the commutator of operators. 
The Hamiltonian $\hat{H}(t)=\hat{H}(\hat{\varrho}(t))$ is implicitly time-dependent, since the system is self-consistently coupled to the Poisson
equation in this one-particle picture.

A mixed state of a quantum system is usually described 
by a positive, Hermitian trace class operator $\hat{\varrho}(t)$, named {\it density matrix 
operator}, acting on 
$L^{2}(\R^{3})$ (see \cite{DaLi90}, \cite{Thi80}).
Hence, $\hat{\varrho}(t)$ is usually represented as an integral operator 
with kernel $\varrho(x,y,t)$:
\begin{eqnarray*}
(\hat{\varrho}(t) f)(x) & = & \int\limits_{\R^{3}} \; \varrho(x,y,t) \; f(y) \; dy \quad 
\forall \; f \in L^{2}(\R^{3}) \; .
\end{eqnarray*}
The Hamiltonian of the Hartree-Fock model can be written as
\begin{equation}
\hat{H}(t) \; = \; \hat{H}_{\MM 0} \; + \; \hat{V}^{\MM H}(t) \; - \; \hat{V}^{\MM HF}(t),
\end{equation}
with the operators
\begin{equation*}
\begin{array}{llll}
\hat{H}_{\MM 0}         & = & - \frac{1}{2} \Delta \qquad  \mbox{ } 
                                & \mbox{free Hamiltonian (without restriction of generality}\\
			&& &	\mbox{we set the Planck constant equal to one)}  , \\[1ex]
\hat{V}^{\MM H}         & = & \hat{V}^{\MM H} [\hat{\varrho}]  & \mbox{Hartree potential} , \\[1ex]
\hat{V}^{\MM HF}        & = & \hat{V}^{\MM HF} [\hat{\varrho}] & \mbox{Hartree-Fock potential}.
\end{array}
\end{equation*}

$\hat{V}^{\MM H}(t)$ is a (local) multiplication operator by
the real valued function 
\begin{eqnarray}\label{v_h_funktion}
V^{\MM H}(x,t) & = & \frac{1}{4 \pi} \; \int\limits_{\R^{3}} \; \frac{n(z,t)}{|x-z|} \; dz \quad . 
\end{eqnarray}
Here, $n(z,t)$ denotes the (real valued) particle density of the system $\hat{\varrho}(t)$ and,
formally, it is obtained by $n(z,t) = \varrho(z,z,t)$. Note that the {\it Hartree potential}
(\ref{v_h_funktion}) is the Newtonian potential solution of the Poisson equation
\begin{equation}\label{poisson_eq}
\triangle V^{\MM H} = - n.
\end{equation}
The (non-local) {\it Hartree-Fock-correction term} (or {\it exchange part}) \linebreak
$[\hat{V}^{\MM HF}(t), \hat{\varrho}]=\hat{V}^{\MM HF}\circ \hat{\varrho}
-\hat{\varrho}\circ \hat{V}^{\MM HF}$ is an integral operator whose first term has the kernel
\begin{eqnarray}
\left( \mbox{kernel} \; ( \hat{V}^{\MM HF} [\hat{\varrho}] \circ \hat{\varrho} ) \right) (x,y,t)
        & = & 
        \frac{1}{4 \pi} \; \int\limits_{\R^{3}} \; 
        \frac{\varrho(x,z,t)}{|x-z|} \; \varrho(z,y,t) \; dz \; , \label{kernel_pres_HF} 
\end{eqnarray}
and hence $\hat{V}^{\MM HF} [\hat{\varrho}]$ has the kernel
\begin{eqnarray}
V^{\MM HF} (x,z,t)  & = &  \frac{1}{4 \pi} \; \frac{\varrho(x,z,t)}{|x-z|} \quad . \label{kernel_pres_HF2}
\end{eqnarray}
Here, we used `$\circ$' to emphasize the composition of operators.

\vspace{3mm}

Often it is convenient to rewrite the initial value problem
(\ref{op_level_initial_value_problem}) as an evolution problem
(integro-differential equation) for the kernel $\varrho$ of $\hat{\varrho}$:
\begin{equation}\label{kern_level_initial_value_problem}
\begin{array}{lll}
\varrho_{t}                & = & \hat{H}_{x} \varrho
                                \; - \; 
                                \hat{H}_{y} \varrho  \; , 
                                \qquad t > 0, \\ [1ex]
\varrho (x,y,t = 0)      & = & \varrho^{\MM I} (x,y),
\end{array}
\end{equation}
with $\varrho^{\MM I}$ denoting the kernel of $\hat{\varrho}^{\MM I}$.
The subscripts $x$ and $y$ indicate that the Hamiltionian $\hat{H}$ acts,
respectively, only on the $x$ or the $y$ variable:
\begin{equation}\label{hamiltonian_variable}
\hat{H}_x (t) = - \frac{1}{2} \Delta_{x} + \hat{V}^{\MM H} (x,t)
          - \hat{V}^{\MM HF}_{x} (t) \quad .
\end{equation}
Here, $\hat{V}^{\MM HF}_{x} (t)$ is the integral operator
(\ref{kernel_pres_HF}) with kernel $V^{\MM HF} (x, z, t)$.
Analogously, $\hat{V}^{\MM HF}_{y} (t)$
has kernel $V^{\MM HF} (z, y, t)$.
The terms $\hat{V}^{\MM H} \circ \hat{\varrho}$ and
$\hat{V}^{\MM HF} \circ \hat{\varrho}$ are obviously quadratic in
$\hat{\varrho}$ which is the main challenge for an existence-uniqueness
analysis of (\ref{op_level_initial_value_problem}) or
(\ref{kern_level_initial_value_problem}). 
On a first glance it would look
easier to analyze the time evolution of the density matrix \textit{function} $\varrho (x, y, t)$
according to (\ref{kern_level_initial_value_problem}) rather than the
evolution of the \textit{operator} $ \hat{\varrho} (t)$. However, the main
problem is to control the ``diagonal'' $n(x,t)$ of $\varrho (x,y,t)$ without
including (unphysically) many spatial derivations into the function space for
$\varrho$. As we shall see in Sec.\ \ref{sec:functional_spaces}, this ``control''
of $n(x,t)$ occurs very naturally when $\hat{\varrho} (t)$ is a trace
class operator (cf.\ \cite{Arn96}, \cite{LiPa} for a more detailed discussion). The
resulting draw-back is that the analysis of the operator evolution equation
(\ref{op_level_initial_value_problem}) is technically much more involved than
analyzing an integro-differential equation of type
(\ref{kern_level_initial_value_problem}).

We remark that there is generally a third approach for analyzing a Hamiltonian
quantum system of form (\ref{op_level_initial_value_problem}). A self-adjoint
trace class operator $\hat{\varrho}$ has a complete orthonormal system
$\{\varphi_j\}_{j\in \mathbb{N}} \subset L^2(\mathbb{R}^3)$ of eigenfunctions
with corresponding eigenvalues $\{\lambda_j\}_{j \in \mathbb{N}} \in
\ell^1(\mathbb{N})$. Its operator kernel then has the ``diagonal''
representation
\begin{equation}\label{g_eigenrepresentation}
\varrho (x, y) = \sum\limits_{j \in \mathbb{N}} \, \lambda_j \varphi_j (x)
\overline{\varphi_j (y)} \quad .
\end{equation}
One easily verifies that, due to the Hamiltonian form of
(\ref{op_level_initial_value_problem}), the eigenvalues $\lambda_j$ of
$\hat{\varrho} (t)$ are constant in time (cf.\ \cite{Mar89}). Hence,
(\ref{op_level_initial_value_problem}) can be rewritten as the following
Schr\"odinger system for the evolution of the eigenfunctions $\varphi_j (x,t)$:
\begin{equation}\label{schroedinger_equation}
\begin{array}{lll}
i \; \frac{\partial}{\partial t} \varphi_j (x, t)
                        & = & \hat{H}_x (t) \varphi_j (x,t), \quad j \in \mathbb{N} \; , 
                                \qquad t > 0 \\ [1ex]
\varphi_j (x, t=0)      & = & \varphi_j^{\MM I} (x), \quad j \in \mathbb{N} \; .
\end{array}
\end{equation}
Here, $\varphi_j^{\MM I}, \; \lambda_j$ are the eigenfunctions and eigenvalues
of $\hat{\varrho}^{\MM I}$. From (\ref{g_eigenrepresentation}) the particle
density becomes
\begin{equation}\label{particle_density_eigenfunc}
n (x,t) = \varrho(x,x,t)=\sum\limits_{j \in \mathbb{N}} \; \lambda_j |\varphi_j (x,t)|^2,
\end{equation}
and hence the Hartree-Fock Hamiltonian reads:
\begin{eqnarray*}
\hat{H}_x (t) \varphi_j (x,t) \!\!\!\!&=&\!\!\!\! - \frac{1}{2} \Delta \varphi_j (x,t) \\
&&\!\!\!\! + \frac{1}{4 \pi}\; \sum\limits_{k \in \mathbb{N}} \; \lambda_k \;
\int\limits_{\mathbb{R}^3} \frac{|\varphi_k (z,t)|^2 \varphi_j (x,t) -
                                 \varphi_k (x,t) \overline{\varphi_k (z,t)}
                                 \varphi_j (z,t)}{|x-z|} \; dz,
\end{eqnarray*}
where the $\lambda_k$ are a-priorly known from the diagonalization of
$\hat{\varrho}^{\MM I}$. In (\ref{schroedinger_equation}) the $\varphi_j$
evolve according to the same Hamiltonian $\hat{H}_x (t)$, and they are only
coupled through the Hartree and Hartree-Fock potential terms.

The formulation (\ref{schroedinger_equation}) of the Hartree-Fock model is
well suited for a mathematical analysis, since (\ref{particle_density_eigenfunc}) yields a
rigorous definition of the particle density in $L^1(\R^3)$. 
Well-posedness of (\ref{schroedinger_equation}) in $H^1 (\mathbb{R}^3)$ and
$H^2 (\mathbb{R}^3)$ was proved for Coulomb interactions in \cite{ChGl75},
and extended to more general interaction potentials in \cite{Gas99}.
\cite{Cast97} analyzes the corresponding Hartree model in
$L^2(\mathbb{R}^3)$, and  the semiclassical limit and large-time behavior of
(\ref{schroedinger_equation}) is investigated in \cite{GIMS98}. Further, quasiperiodic solutions to the 
Hartree-Fock system in an external electro-magnetic field are constructed in \cite{DIL}.

We point out that (\ref{schroedinger_equation}) is \textit{almost equivalent}
to (\ref{op_level_initial_value_problem}): the unique solution of
(\ref{schroedinger_equation}) also solves
(\ref{op_level_initial_value_problem}), but uniqueness of the solution
$\hat{\varrho} (t)$ of (\ref{op_level_initial_value_problem}) does not follow.
Hence we shall directly analyze (\ref{op_level_initial_value_problem}) in this
paper, following the approach of Ref.\ \cite{Arn96}. This strategy was also
used in \cite{BoDPFa74} to prove existence and uniqueness of a trace class
operator-solution to the Hartree-Fock system. In that paper, however, the
Coulombian two-particle interaction potential $|x|^{-1}$ (appearing in (\ref{v_h_funktion}) and
(\ref{hamiltonian_variable})) was approximated by a bounded function which greatly simplified the
analysis (The Hartree-Fock term is then locally Lipschitz in the space of trace class operators, and
the kinetic energy is not needed). 
A second motivation for our approach is that the reformulation
(\ref{schroedinger_equation}) becomes impossible for open quantum systems,
since the eigenvalues $\lambda_j$ would then be time-dependent. Open quantum
systems are important in many fields of applications (quantum diffusion,
coupling to a heat bath, cf.\
\cite{AlFa}, \cite {Da}, \cite{ArSp03}, and references
therein) and they are modelled by augmenting the right hand side of
(\ref{op_level_initial_value_problem}) by interaction terms of Lindblad form (cf.\ \cite{Arn97},
\cite{Li}).
In our subsequent analysis we shall not include such (bounded) Lindblad
operators, but they would not pose any additional analytical problems.

The paper is organized as follows: in Sec.\ \ref{sec:functional_spaces} we introduce the functional
setting for our subsequent analysis. Using perturbation techniques from semigroup theory we prove
the existence of local-in-time solutions for
(\ref{op_level_initial_value_problem}) in Sec.\ 
\ref{sec:local_time_solution}. In Sec.\  \ref{sec:global_time_solution} we give a rigorous proof that 
our solutions to the Hartree-Fock system are mass and energy conserving. 
These a-priori estimates then imply that the constructed solutions are global.

\renewcommand{\thesection}{\Roman{section}}
\section{Notations and functional setting}\label{sec:functional_spaces}
\renewcommand{\thesection}{\arabic{section}}

We shall use the notation ${\cal I}_{\MM 1}$ and ${\cal I}_{\MM 2}$ for the spaces of, resp., trace
class operators and Hilbert-Schmidt operators acting on $L^{2} (\R^3)$. They are equipped with the norms
(see also \cite{RSI})
\begin{eqnarray}\label{kern_hartree_n2} 
  \3norm \hat{\varrho} \3norm_{1} & := &  
  \mbox{Tr} |\hat{\varrho}| \; , \\ \3norm \hat{\varrho} \3norm_{2} 
  & := &  ( \mbox{Tr} |\hat{\varrho}|^{ 2})^{1/2} \; . 
\end{eqnarray}

Since
$\hat{H}_{\MM 0} = - \Delta/2 \geq 0\;$, the operator $\hat{H}_{\MM 0}^{1/2} := \sqrt{\hat{H}}_{\MM 0}$
is well-defined (having the symbol $|\xi|/\sqrt{2}$ in Fourier space) and we can introduce the normed linear spaces
\begin{equation}\label{Z_def}
\begin{array}{lll}
Z 
        & := &
        \{ \hat{\varrho} \in {\cal I}_{\MM 1} \; | \; \hat{\varrho} \; \mbox{is Hermitian}, \; 
        \overline{\hat{H}_{\MM 0}^{1/2} \hat{\varrho} \hat{H}_{\MM 0}^{1/2}} 
        \in {\cal I}_{ 1}  \},\\ [1ex]
\| \hat{\varrho} \|_{ Z}
        & := & \3norm \hat{\varrho} \3norm_{ 1}
        \; + \; 
        \3norm \overline{\hat{H}_{\MM 0}^{1/2} \hat{\varrho} \hat{H}_{\MM 0}^{1/2}} \3norm_{ 1},
\end{array}
\end{equation}
and 
\begin{equation}\label{Y_def}
\begin{array}{lll}
Y 
        & := &
        \{ \hat{\varrho} \in {\cal I}_{\MM 1} \; | \; \hat{\varrho} \; \mbox{is Hermitian}, \; 
        \hat{H}_{\MM 0} \hat{\varrho}
        \in {\cal I}_{ 1}  \}, \\[1ex]
\| \hat{\varrho} \|_{ Y}
        & := & \3norm \hat{\varrho} \3norm_{ 1}
        \; + \; 
        \3norm \hat{H}_{\MM 0} \hat{\varrho} \3norm_{ 1}     \qquad .
\end{array}
\end{equation}
Here, 
$\overline{\hat{H}_{\MM 0}^{1/2} \hat{\varrho} \hat{H}_{\MM 0}^{1/2}}$ denotes the
closure (on $L^2 (\mathbb{R}^3)$) of the operator
$\hat{H}_{\MM 0}^{1/2} \hat{\varrho}\hat{H}_{\MM 0}^{1/2}$,
which is only defined on (a subset of) $H^1 (\mathbb{R}^3)$.
Due to the compactness of $\hat{\varrho} \in Z$, however, it can be extended
to a bounded operator on $L^2 (\mathbb{R}^3)$. In the sequel we shall mostly
suppress this closure symbols to keep the notation simple.
In Lemma \ref{Y_in_Z} we shall show that $Y \subset Z$ holds .

We shall denote operators in the form $\hat{\varrho}$ (with an overwritten``hat'') to
distinguish them from their kernels.
For a self-adjoint operator $\hat{\varrho} \in {\cal I}_{\MM 2}$ its kernel
$\varrho$ has the $diagonal$ representation 
(\ref{g_eigenrepresentation}), 
where $\{ \lambda_{j}\}_{j \in \N} \in l^{2}(\N)$ and the complete orthonormal system 
$\{ \varphi_{j}\}_{j \in \N} \subset L^{2}(\R^{3})$
are, resp., the eigenvalues and eigenfunctions of $\hat{\varrho}$.
$\hat{\varrho}$ is Hermitian if and only if its kernel satisfies
$\varrho(x,y) = \overline{\varrho(y,x)}$
(the bar denotes complex conjugation).
For self-adjoint operators 
$\hat{\varrho} \in {\cal I}_1$ we even have $\{ \lambda_{j}\}_{j \in \N} \in l^{1}(\N)$. 
Recalling the definition of the
{\it particle density} in (\ref{particle_density_eigenfunc}), it is now possible to estimate $n(x)$
via the trace norm of $\hat{\varrho}$:
\begin{equation*}
\|n\|_{L^1(\R^3)} \le \3norm\hat{\varrho}\3norm_1 = \sum \limits_{j \in \N} \; |\lambda_{j}| \,,
\end{equation*}
with an equality for $\hat{\varrho}$ positive. This natural control of the $L^1$-norm of $n$
constitutes the main justification for considering the Hartree-Fock system
(\ref{op_level_initial_value_problem}) in ${\cal I}_{\MM 1}$ instead of the PDE
(\ref{kern_level_initial_value_problem}).

While physical quantum states only lie in the cone of positive operators
of $Z$ or $Y$, we shall consider here the whole spaces as this simplifies 
the subsequent analysis. From a physical point of view the space $Z$ comprises quantum states with finite mass (i.e.\ Tr$\hat{\varrho} <
\infty$) and finite {\it kinetic energy}, which
is defined as
\begin{eqnarray}{\label{def_E_kin}}
E_{kin} (\hat{\varrho}) 
        & := &
        \mbox{Tr} (\hat{H}_{\MM 0}^{1/2} \; \hat{\varrho} \; \hat{H}_{\MM 0}^{1/2}) \;.
\end{eqnarray}
For $\hat{\varrho} \geq 0$ this equals 
$\3norm \hat{H}_{\MM 0}^{1/2}  \hat{\varrho}  \hat{H}_{\MM 0}^{1/2} \3norm_{ 1}$.
Moreover, we can compute the kinetic energy ({\ref{def_E_kin}}) in terms of the eigenvector
decomposition of $\hat{\varrho}$ as
\begin{eqnarray}\label{Ekin_typischI}
E_{kin} (|\hat{\varrho}|) 
         = 
        \frac{1}{2}
        \sum\limits_{j \in \N} \; |\lambda_{j}| \; \| \nabla \varphi_{j}(x) \|^{2}_{L^{2}(\R^3)}
         = 
        \3norm \hat{H}_{\MM 0}^{1/2} \hat{\varrho} \hat{H}_{\MM 0}^{1/2} \3norm_{1} 
\end{eqnarray}
(cf.\ Ref.\ \cite{Arn96}, Lemma A.1). Hence, $\hat{\varrho}\in Z$ implies
\begin{eqnarray}\label{varphi_in_H1}
\{ \varphi_{j}\}_{j \in \N}
        & \subset &
        H^{\MM 1}(\R^{3}),
\end{eqnarray}
and as a consequence
\begin{equation}
\begin{array}{lll} 
\sum\limits_{j \in \N} \; | \lambda_{j} | \; \big\| \varphi_{j}(x) \big\|^{2}_{H^{1}(\R^3)}
        \!\!\!& = &\!\!\!
\sum\limits_{j \in \N} \; | \lambda_{j} | \; 
        \big\| \varphi_{j}(x) \big\|^{2}_{L^{2}(\R^3)} 
        \; + \; 
        \sum\limits_{j \in \N} \; | \lambda_{j} | \; 
        \big\| \nabla \varphi_{j}(x) \big\|^{2}_{L^{2}(\R^3)} \label{rep_eigen_mit_1norm} \\[1ex]
        \!\!\!& = &\!\!\!
        \3norm \hat{\varrho} \3norm_{1}  
        \; + \; 
        2 \3norm \hat{H}_{\MM 0}^{1/2} \; \hat{\varrho} \; \hat{H}_{\MM 0}^{1/2} \3norm_{1} \; ,
\end{array}
\end{equation}
which is equivalent to the $Z$-norm.

As we shall see in the next section, initial conditions in the spaces  $Z$ and $Y$ give rise to,
resp., mild and classical solutions of (\ref{op_level_initial_value_problem}).

Some properties of $Z$ and $Y$ are stated in 

\begin{lemma} \mbox{ } \label{Y_in_Z} \\
\emph{(a)} $Z$ is a real Banach space.\\
\emph{(b)} $Y$ is a real Banach space.\\
\emph{(c)} $Y$ is a dense subspace of $Z$.
\end{lemma}

\proof (a)  The linear subspace of Hermitian trace class operators is
closed in ${\cal I}_{\MM 1}$.
Now let 
$\{ \hat{\varrho}_{j} \}_{j \in \N}$ be a Cauchy sequence
in $Z$. Hence $\exists \; \hat{\sigma},\hat{\gamma}\in {\cal I}_{\MM 1}$ such that, for $j \rightarrow \infty$,
\begin{eqnarray}{\label{kern2_seq}}
\hat{\varrho}_{j} \longrightarrow \hat{\sigma},  \quad \quad
\hat{H}_{\MM 0}^{1/2}  \hat{\varrho}_j \hat{H}_{\MM 0}^{1/2}  \longrightarrow \hat{\gamma} \quad \mbox{in } {\cal
I}_{\MM 1},
\end{eqnarray}
and the corresponding Cauchy sequences of the
kernels satisfy, as $j \rightarrow \infty$:
\begin{eqnarray*}
\varrho_{j}(x,y) \longrightarrow \sigma(x,y) \; , \quad
\hat{H}_{{\MM 0},x}^{1/2} \hat{H}_{{\MM 0},y}^{1/2} \varrho_j(x,y)  \longrightarrow \gamma(x,y) \quad \mbox{in}\; L^2(\R^6)\;.
\end{eqnarray*}
By Fourier transforming in both $x$ and $y$ (and denoting its dual variables by $\xi$ and $\eta$)
we have
\begin{eqnarray*}
{\cal F}\varrho_j(\xi,\eta)   &\longrightarrow& {\cal F}\sigma (\xi,\eta) \quad
\mbox{in} \; L^2 \big(\R^3_{\xi}\times \R^3_{\eta}, (1 + \frac{1}{4}|\xi |^2|\eta |^2) d\xi d\eta \big) \;,
\end{eqnarray*}
which is a complete space. Therefore 
\begin{eqnarray*}
\gamma(x,y)   &=& \hat{H}_{{\MM 0},x}^{1/2} \hat{H}_{{\MM 0},y}^{1/2} \sigma (x,y) \quad
\mbox{ and } \quad \hat{\varrho}_j \longrightarrow \hat{\sigma} \quad \mbox{ in }Z.
\end{eqnarray*}

(b)  Same argument as for $(a)$ (see Ref.\ \cite{Arn96} for details).\\

(c)  The proof is divided into two steps: first we define an auxiliary space
$X\subset Y$, then we prove that $X$ is dense in $Z$. Let
\begin{eqnarray}
X 
        & := &
        \{ \hat{\varrho} \in {\cal I}_{\MM 1} \; | \; \hat{\varrho} \; \mbox{Hermitian}, \; 
        \Delta \hat{\varrho} \Delta
        \in {\cal I}_{\MM 1}  \} ,\\[1ex] \label{norm_X}
\| \hat{\varrho} \|_{X}
        & := & \3norm \hat{\varrho} \3norm_{1}
        \; + \; 
        \3norm \Delta \hat{\varrho} \Delta \3norm_{1}     \qquad .
\end{eqnarray}

$Step$ $1$: show $X \subseteq Y$.\\
Let $\hat{\varrho} \in X$ and decompose it (and at the same time $\Delta \hat{\varrho} \Delta$) in its positive
and negative parts: $\hat{\varrho}  =  \hat{\varrho}_{+} - \hat{\varrho}_{-}$,\; 
$\Delta \hat{\varrho} \Delta =  \Delta \hat{\varrho}_{+} \Delta - \Delta \hat{\varrho}_{-} \Delta$ \;
with \;
$\hat{\varrho}_{\pm} \geq 0, \Delta \hat{\varrho}_{\pm} \Delta \geq 0 $.
From $\hat{\varrho}_{\pm}, \Delta \hat{\varrho}_{\pm} \Delta \in {\cal I}_{1}$ follows $\sqrt{\hat{\varrho}_{\pm}}, \Delta
\sqrt{\hat{\varrho}_{\pm}} \in {\cal I}_{2}$. Hence $\Delta \hat{\varrho}_{\pm} \in {\cal I}_{1}$, which implies $\hat{\varrho} \in Y$.

$Step$ $2$: show that $X$ is dense in $Z$.\\
Let $\hat{\varrho} \in Z$. Assume without restriction of generality that $\hat{\varrho} \geq 0$ (otherwise
separate into $\hat{\varrho}_{\pm}$). We have
\begin{eqnarray*}
\|\hat{\varrho}\|_{Z} = \sum_{j=1}^{\infty}  \lambda_j \Big(\| \varphi_j \|^{2}_{L^2(\R^3)} + \frac{1}{2}\|
\nabla \varphi_j \|^{2}_{L^2(\R^3)} \Big),
\end{eqnarray*}
where $\lambda_j$, $\varphi_j$ are the eigenvalues and
eigenfunctions of $\hat{\varrho}$.
For all $\epsilon > 0 : \exists N\in\N$ with $\hat{\varrho}^N$
satisfying \;$\|\hat{\varrho} - \hat{\varrho}^N \|_{Z} < \epsilon/2$, with the kernel $\varrho^N(x,y)=
\sum_{j=1}^{N}  \lambda_j  \varphi_j(x) \overline{\varphi_j(y)}$.
For each $n \in \N$ we consider approximations $\hat{\sigma}_n$, which we define in terms of their
kernels:
\begin{equation*}
\sigma_n(x,y) = \sum_{j=1}^{N}  \lambda_j  \psi_j^n(x)
\overline{\psi_j^n(y)} ,
\end{equation*}
with appropriate functions $\psi_j^n \in H^2(\R^3)$ such that  $\psi_j^n \stackrel{n \rightarrow
\infty}{\longrightarrow} \varphi_j$ in $H^1(\R^3)$. Note, that we do not require 
$\{\psi_j^n\}_{j=1}^N$ to be orthonormal. We have $\hat{\sigma}_n \in X$ (since
$\psi_j^n \in H^2(\R^3)$
and rank$(\hat{\sigma}_n)\leq N$), $\hat{\sigma}_n \ge 0$ as sum of positive operators. Further,
\begin{eqnarray*}
\hat{\sigma}_n \stackrel{n \rightarrow \infty}{\longrightarrow} \hat{\varrho}^N \quad \mbox{and} \quad 
\hat{H}_{\MM 0}^{1/2} \hat{\sigma}_n \hat{H}_{\MM 0}^{1/2} \; \stackrel{n \rightarrow
\infty}{\longrightarrow}\;
\hat{H}_{\MM 0}^{1/2} \hat{\varrho}^N \hat{H}_{\MM 0}^{1/2} 
\end{eqnarray*}
in the strong operator topology, since we have $\forall f \in L^2(\R^3)$:
\begin{eqnarray*}
(\hat{\sigma}_n f)(x) &=& \sum_{j=1}^N \lambda_j \psi_j^n(x)
 \int_{\R^3} f(y) \overline{\psi_j^n(y)} dy \\
&\stackrel{\mbox{\scriptsize $n \rightarrow \infty$}}{\longrightarrow} &
\sum_{j=1}^N \lambda_j \varphi_j(x)
\int_{\R^3} f(y) \overline{\varphi_j(y)} dy = (\hat{\varrho}^N f)(x) \; \mbox{in $L^2(\R^3)$},\\
(\hat{H}_{\MM 0}^{1/2} \hat{\sigma}_n \hat{H}_{\MM 0}^{1/2} f)(x) &=&
\sum_{j=1}^N \lambda_j \hat{H}_{\MM 0}^{1/2} \psi_j^n(x) \int_{\R^3} f(y) \hat{H}_{\MM 0}^{1/2}
\overline{\psi_j^n(y)} dy \\
&\stackrel{\mbox{\scriptsize $n \rightarrow \infty$}}{\longrightarrow} &
(\hat{H}_{\MM 0}^{1/2} \hat{\varrho}^N \hat{H}_{\MM 0}^{1/2} f)(x)  \; \mbox{in $L^2(\R^3)$}.
\end{eqnarray*}
Their norms satisfy
\begin{eqnarray*}
\3norm \hat{\sigma}_n \3norm_1 = \mbox{Tr}\hat{\sigma}_n =\sum_{j=1}^N \lambda_j \| \psi_j^n
\|^2_{L^2(\R^3)} \quad \stackrel{n \rightarrow \infty}{\longrightarrow} \quad
\sum_{j=1}^N \lambda_j \| \varphi_j \|^2_{L^2(\R^3)} = \3norm \hat{\varrho}^N \3norm_1 ,\\
\3norm \hat{H}_{\MM 0}^{1/2} \hat{\sigma}_n \hat{H}_{\MM 0}^{1/2} \3norm_1 =
\frac{1}{2} \sum_{j=1}^N \lambda_j \| \nabla \psi_j^n
\|^2_{L^2(\R^3)} \qquad\qquad\qquad\qquad\qquad \\ \stackrel{n \rightarrow \infty}{\longrightarrow} \quad
\frac{1}{2} \sum_{j=1}^N \lambda_j \| \nabla \varphi_j \|^2_{L^2(\R^3)} = \3norm \hat{H}_{\MM 0}^{1/2}
\hat{\varrho}^N \hat{H}_{\MM 0}^{1/2} \3norm_1 .
\end{eqnarray*}
The two equalities on the left hand side are easily verified by evaluating the trace in an arbitrary orthonormal system of $\L^2(\R^3)$.
By Gr\"umm's convergence theorem (\cite{Sim78}, Theorem 2.19) we obtain 
\begin{eqnarray*}
\3norm \hat{\sigma}_n - \hat{\varrho}^N \3norm_1 \stackrel{n \rightarrow \infty}{\longrightarrow} 0,
\quad \3norm \hat{H}_{\MM 0}^{1/2} \hat{\sigma}_n \hat{H}_{\MM 0}^{1/2} - \hat{H}_{\MM
0}^{1/2} \hat{\varrho}^N \hat{H}_{\MM 0}^{1/2} \3norm_1 \stackrel{n \rightarrow
\infty}{\longrightarrow} 0 \quad.
\end{eqnarray*}
Since $\hat{\sigma}_n \stackrel{n \rightarrow \infty}{\longrightarrow} \hat{\varrho}^N$ in $Z$, 
$X$ is dense in $Z$. And so is $Y$.
\qed
\vspace{5mm}


\renewcommand{\thesection}{\Roman{section}}
\section{Local-In-Time Solution}\label{sec:local_time_solution}
\renewcommand{\thesection}{\arabic{section}}

This section is concerned with the local-in-time solution of the Hartree-Fock system ({\ref{op_level_initial_value_problem}}).
We rewrite ({\ref{op_level_initial_value_problem}}) as
%
\begin{equation}\label{HF_system}
\begin{array}{lll}
\hat{\varrho}_{t} (t)      
        & = &    -i \;[\hat{H}_{\MM 0}, \hat{\varrho} (t)]
        \; - \;  i \;
         [\hat{V}^{\MM H}(t), \hat{\varrho} (t)]
        \; + \;  i \;
         [\hat{V}^{\MM HF}(t), \hat{\varrho} (t)]\\[1ex]
        & = & h_{\MM 0} (\hat{\varrho}) \; + \; F (\hat{\varrho}) \; ,  \\ [1ex]
\hat{\varrho} (0) & = & \hat{\varrho}^{\MM I} \; ,               
\end{array}
\end{equation}
where we formally define the operators $h_{\MM 0}$ and $F$ as: 
\begin{equation}\label{def_h0_F} 
\begin{array}{lll}
h_{\MM 0} (\hat{\varrho})       
        & := &
        -i \; [\hat{H}_{\MM 0}, \hat{\varrho} (t)] \; ,\\[1ex]
F (\hat{\varrho})
        & := &
        -i \; [\hat{V}^{\MM H}(t) \; - \; \hat{V}^{\MM HF}(t), \hat{\varrho} (t)] \;.
\end{array}
\end{equation}

%
%
First we consider the free evolution equation
\begin{equation}\label{free_evol}
\begin{array}{lll}
\hat{\varrho}_{t} (t) & = & h_{\MM 0} (\hat{\varrho}) \; , \; t>0, \\[1 ex]
\hat{\varrho} ( 0 )   & = & \hat{\varrho}^{\MM I} 
\end{array}
\end{equation}
in the ``energy spaces" $Z$ and $Y$. 
This linear problem admits a unique global solution in ${\cal I}_1$ (cf.\
\cite{DaLi92} Chap.\ 5). It can be represented via 
the isometric $C_0-$evolution group $\{ G_{\MM 0} (t) , t \in \R \}$, which reads
\begin{eqnarray}\label{rep_isom_evo_group}
G_{\MM 0} (t) \hat{\varrho}
        & = &
        e^{-i \hat{H}_{\MM 0} t} \; \hat{\varrho} \; e^{i \hat{H}_{\MM 0} t} \; .
\end{eqnarray}
Its {\it infinitesimal generator} $h_{\MM 0}$ is defined as
\begin{eqnarray*}
{\cal D}(h_{\MM 0}) &=& \{ \hat{\varrho} \in {\cal I}_1 \;| \; \hat{\varrho}{\cal D}(\hat{H}_{\MM 0}) \subset {\cal D}(\hat{H}_{\MM
0})=H^2(\R^3), \;
 (\hat{H}_{\MM 0}\hat{\varrho} - \hat{\varrho}\hat{H}_{\MM 0}) \; \mbox{is an operator}\\
 && \mbox{with domain $H^2(\R^3)$ with an $L^2(\R^3)$ extension }
 \; \overline{\hat{H}_{\MM 0}\hat{\varrho} - \hat{\varrho}\hat{H}_{\MM 0}} \in  
 {\cal I}_1\}\;,\\
h_{\MM 0} (\hat{\varrho})       
        & = &
        -i \; \overline{[\hat{H}_{\MM 0}, \hat{\varrho}]}. 
\end{eqnarray*}
We remark that $Y \subset {\cal D}(h_{\MM 0})$, while $Z \not\subseteq {\cal D}(h_{\MM
0})$.
The group $\{ G_{\MM 0} (t),\; t \in \R \}$ preserves Hermiticity and positivity in ${\cal I}_1$(cf.\
Ref.\ \cite{DaLi92} Theorem 5.1).

The following Theorem \ref{G_0_restic} states that the restriction of $\{ G_{\MM 0} (t) , t \in \R \}$
to $Z$ and $Y$ yields a global solution of ({\ref{free_evol}}) in these two spaces.\\
\begin{theorem}\label{G_0_restic}
The evolution group $G_{\MM 0}$ restricted to $Z$ (resp. $Y$) is 
an isometric $C_{\MM 0}-$evolution group on $Z$ (resp. $Y$).
\end{theorem}

\proof 
%
Since $\hat{H}_{\MM 0}^{1/2}$ and $G_{\MM 0} (t)$ commute,
it follows directly from the corresponding properties of $G_{\MM 0} (t)$ on ${\cal I}_{\MM 1}$.
%
\qed

\vspace{3mm}
As a second step we shall consider (\ref{HF_system}) as a perturbation of ({\ref{free_evol}}) by the {\it perturbation}
operator $F$ (see \cite{Paz83} Theorem 6.1.4).
We shall demonstrate that $F$ is locally Lipschitz in $Z$ and in $Y$, which guarantees the existence of a unique
local-in-time solution for (\ref{HF_system}), as stated in Theorem \ref{th_loc_sol}. \\
Next we show that the perturbation $F$ is locally Lipschitz. It is demonstrated by Corollary \ref{theore_F_lipsch}, which is the result of the following three lemmata
where we prove that $F$ maps $Z$ into $Z$, and resp., $Y$ into $Y$.

In the sequel $C$ denotes generic, but not necessarily equal constants. $L^p_w$ will denote the
weak $L^{p}-$space (cf.\ \cite{RSII}, e.g.).


\begin{lemma}\label{VHF_bounded}
Let $\hat{\varrho} \in Z$. Then
$\hat{V}^{\MM HF} [\hat{\varrho}]\in {\cal B}(L^{2}(\R^{3}))$.
\end{lemma}

\proof
The eigenfunctions of $\hat{\varrho}$ satisfy (cf.\ ({\ref{g_eigenrepresentation}}),
({\ref{varphi_in_H1}}))
$\varphi_{j} \in H^{\MM 1}(\R^{3})$, which is continuously embedded in $L^{p}(\R^{3})$ 
for $2 \leq p \leq 6$. For arbitrary $f \in L^2(\R^3)$ we use
$\varphi_{j} f \in L^{p}(\R^{3})$ for $ 1 \leq p \leq 3/2$ and $1/|x| \in L^{3}_{w}(\R^{3})$. Then the
generalized Young inequality yields (where `$*$' denotes the convolution operator):
\begin{eqnarray}
(\varphi_{j} \; f) \ast \frac{1}{|x|}
        & \in &
        L^{q}(\R^{3}) \; , \quad 3 < q < \infty \; , \label{use_gen_young}
\end{eqnarray}
and the H\"older inequality implies:
\begin{eqnarray}
\varphi_{j} \; \left( (\varphi_{j} f) \ast \frac{1}{|x|} \right)
        & \in &
        L^{p}(\R^{3}) \; , \quad \frac{6}{5} < p < 6 \quad .\; 
\end{eqnarray}
Using the kernel ({\ref{kernel_pres_HF2}}) of $\hat{V}^{\MM HF}$ we estimate for all $f \in L^{2}(\R^{3})$:
\begin{eqnarray*}
4 \pi \|\hat{V}^{\MM HF}[\hat{\varrho}] f\|_{L^{2}(\R^{3})} 
        &= &   \Big\| \; \int\limits_{\R^{3}} \; 
        \frac{\varrho(x,z)}{|x-z|} \; f(z) \; dz \; \Big\|_{L^{2}(\R^{3}_x)} \\
        & \leq &
        \sum\limits_{j \in \N} \; | \lambda_{j} | \; 
        \Big\| \varphi_{j}(x)
        \int\limits_{\R^{3}} \; 
        \frac{ \overline{\varphi_{j}(z)} }{|x-z|} \; f(z) \; dz \; \Big\|_{L^{2}(\R^{3}_x)} \\
        & \leq &
        \sum\limits_{j \in \N} \; | \lambda_{j} | \; 
        \big\| \varphi_{j} \big\|_{L^{3}(\R^{3})} \;
        \Big\| \big( \overline{\varphi_{j}(x)} f(x) \big) \ast \frac{1}{|x|} \Big\|_{L^{6}(\R^{3})} \nonumber \\
        & \leq &
        \sum\limits_{j \in \N} \; | \lambda_{j} | \;
        \big\| \varphi_{j} \big\|_{L^{3}(\R^{3})} \;
        \big\| \varphi_{j} f \big\|_{L^{6/5}(\R^{3})} \; \Big\| \frac{1}{|x|} \Big\|_{L^{3}_{w}(\R^{3})} \nonumber \\
        & \leq &
        C \;
        \Big( \sum\limits_{j \in \N} \; | \lambda_{j} | \;
        \big\| \varphi_{j}(x) \big\|^{2}_{H^{1}(\R^{3})} \Big) \;
        \big\| f \big\|_{L^{2}(\R^{3})} \nonumber \\
        & \leq &
        C \; \| \varrho \|_{Z} \;  \big\| f \big\|_{L^{2}(\R^{3})} \; ,
\end{eqnarray*}
where we used (\ref{rep_eigen_mit_1norm}) for the last estimate.
Hence, $\|\hat{V}^{\MM HF}[\hat{\varrho}] \|_{{\cal B}(L^{2}(\R^{3}))} \le C \| \hat{\varrho}
\|_{Z}$.
\qed
\vspace{3mm}

\begin{lemma}\label{lemma_in_I}
\emph{(a)} Let $\hat{\varrho} \in Z$.
Then $\hat{H}_{\MM 0}^{1/2} \; \hat{V}^{\MM HF} \circ \hat{\varrho} \; \hat{H}_{\MM 0}^{1/2} \in {\cal I}_{\MM 1}$.\\
\emph{(b)} Let $\hat{\varrho} \in Y$.
Then $\hat{H}_{\MM 0} \; \hat{V}^{\MM HF} \circ \hat{\varrho} \in {\cal I}_{\MM 1}$.
\end{lemma}

\proof
For the subsequent analysis we first note that 
$\hat{H}_{\MM 0}^{1/2}$ and $\nabla$
are ``equivalent operators'' in the sense that 
\begin{equation}\label{def_K_G}
\hat{H}_{\MM 0}^{1/2}
        \;  = \;
        \sum_{j=1}^{3} \; K_{j} \; \partial_{x_{j}}
        \quad 
        \mbox{and} 
        \quad 
        \partial_{x_{j}} \; = \; - \sqrt{2} \; \hat{H}_{\MM 0}^{1/2} \; R_{j} \; .
\end{equation}
Here $K_{j}$ and $R_{j}\; (j=1,2,3)$ are bounded pseudo-differential operators defined by their symbols (in Fourier space) 
\begin{equation}
k_{j}(\xi) \;  = \;  \frac{i}{\sqrt{2}} \; 
        \frac{|\xi| \; \mbox{sgn} \; \xi_{j}}{|\xi_{\MM 1}| + |\xi_{\MM 2}| + |\xi_{\MM 3}| } 
        \quad 
        \mbox{and} 
        \quad 
        r_{j}(\xi) \;  = \; i \; \frac{\xi_{j}}{|\xi|} \; .
\end{equation}
$R_{j}$ denotes the {\it j-th Riesz transform} operator in $\R^3$ (cf.\ \cite{Ste70}, p. 58 for definition
and details).  
Since $k_j,r_j \in L^\infty(\R^3)$ we have $K_j,R_j \in {\cal B}(L^2(\R^3))$, and hence:
\begin{eqnarray*}
\hat{H}_{\MM 0}^{1/2} \; \hat{\varrho} \; \hat{H}_{\MM 0}^{1/2} \; \in {\cal I}_{\MM 1}
        & \mbox{ if and only if } &
        \partial_{x_{j}} \; \hat{\varrho} \; \partial_{x_{k}} \; \in {\cal I}_{\MM 1} \quad \forall j,k = 1,2,3 \; .
\end{eqnarray*}
\\
$Part$ (a): To prove the assertion we will show
\begin{equation*}
\partial_{x_{k}} \; \hat{V}^{\MM HF} \circ \hat{\varrho} \; \partial_{x_{j}} \; 
        \in {\cal I}_{\MM 1} \quad \forall \; j,k = 1,2,3 \; .
\end{equation*}
Since $\hat{V}^{\MM HF}$ is an integral operator
with kernel (\ref{kernel_pres_HF2}),
$\partial_{x_{k}} \,\hat{V}^{\MM HF}$ is also
an integral operator with the kernel
\begin{eqnarray*}
\partial_{x_{k}} \; V^{\MM HF} (x,y)
        & = &
        \frac{1}{4 \pi} \; \frac{\partial_{x_{k}} \; \varrho(x,y)}{|x-y|}
        \; - \; 
        \frac{1}{4 \pi} \; \frac{\varrho(x,y)}{|x-y|^{3}} \; (x_{k} - y_{k})
        \; .
\end{eqnarray*}
The last two terms define, resp., the functions $D^{k}_{\MM 1}(x,y)$ and $-D^{k}_{\MM 2}(x,y)$.
For the two integral operators $\hat{D}^{k}_{\MM 1}$ and $\hat{D}^{k}_{\MM 2}$
belonging to these kernels we will now show:
\begin{eqnarray}\label{D1_in_I1}
\hat{D}^{k}_{\MM 1} \; \hat{\varrho} \; \partial_{x_{j}}
        & \in & {\cal I}_{\MM 1}   
        \qquad \forall \; j,k \;  = 1,2,3 \; ,  \\[1ex]\label{D2_in_I1}
        \hat{D}^{k}_{\MM 2} \; \hat{\varrho} \; \partial_{x_{j}}
        & \in & {\cal I}_{\MM 1} 
        \qquad \forall \; j,k \;  = 1,2,3 \; ,
\end{eqnarray}
which will then imply
\begin{equation*}
\partial_{x_k} \; \hat{V}^{\MM HF} \; \circ \hat{\varrho} \; \partial_{x_{j}} 
        \; = \; 
        \hat{D}^{k}_{\MM 1} \; \hat{\varrho} \; \partial_{x_{j}}
        \; + \; 
        \hat{D}^{k}_{\MM 2} \; \hat{\varrho} \; \partial_{x_{j}}
        \; \in \; 
        {\cal I}_{\MM 1}  \qquad
        \forall \; j,k \;  = 1,2,3 \; .
\end{equation*}

To prove (\ref{D1_in_I1}) we consider the identity 
\begin{eqnarray}\label{Dk_decomp}
\hat{D}^{k}_{\MM 1} \; \hat{\varrho} \; \partial_{x_{j}}
        & = &
        \hat{D}^{k}_{\MM 1} \; \big( \hat{H}_{\MM 0}^{1/2} + I \big)^{-1} \; 
        \big( \hat{H}_{\MM 0}^{1/2} \; \hat{\varrho} \; \partial_{x_{j}} \; + \; \hat{\varrho} \;
        \partial_{x_{j}} \big) \;  .
\end{eqnarray}
In (\ref{Dk_decomp}), $\hat{H}_{\MM 0}^{1/2} \; \hat{\varrho} \; \partial_{x_{j}} \in {\cal I}_{\MM 1}$ 
follows with (\ref{def_K_G}) immediately from $\hat{\varrho} \in Z$.
To prove
$\hat{\varrho} \; \partial_{x_{j}} \in {\cal I}_{\MM 1}$
we decompose the self-adjoint operator $\hat{\varrho}$
into its positive and its negative parts:
$\hat{\varrho} \; = \; \hat{\varrho}_{+} \; - \; \hat{\varrho}_{-}
\; \mbox{with} \; \hat{\varrho}_{\pm} \geq 0$.
Hence we have
\begin{equation*}
\hat{\varrho}_{\pm} \; \hat{H}_{\MM 0}^{1/2} 
        \; = \; 
        \hat{\varrho}_{\pm}^{1/2} \;
        \Big ( \hat{\varrho}_{\pm}^{1/2} \hat{H}_{\MM 0}^{1/2} \Big) 
        \; \in \; {\cal I}_{\MM 1} \; ,
\end{equation*}
since both factors of the right hand side are in ${\cal I}_{\MM 2}$ (for the second
factor we have 
$\hat{H}_{\MM 0}^{1/2} \hat{\varrho}_{\pm} \hat{H}_{\MM 0}^{1/2} \in {\cal I}_{\MM 1}$
if and only if
$\hat{\varrho}_{\pm}^{1/2} \hat{H}_{\MM 0}^{1/2} \in {\cal I}_{\MM 2}$, cf.\ Ref.\ \cite{Arn96}, 
Lemma A.1). (\ref{def_K_G}) then yields $\hat{\varrho} \;
\partial_{x_{j}} \in {\cal I}_{\MM 1}$ and it remains to show that
$\hat{D}^{k}_{\MM 1} \; \big( \hat{H}_{\MM 0}^{1/2} + I \big)^{-1} \in {\cal B}(L^{2}(\R^{3}))$.
Since
$\big( \hat{H}_{\MM 0}^{1/2} + I \big)^{-1}$ is a bounded operator from
$L^2 (\mathbb{R}^3)$ to $H^1 (\mathbb{R}^3)$  we set for any
$f \in L^2 (\mathbb{R}^3):\;$
$\big( \hat{H}_{\MM 0}^{1/2} + I \big)^{-1} f =: g \in H^{1}(\R^{3})$.
Hence it suffices to prove
\begin{equation*}
\| \hat{D}^{k}_{\MM 1} \; g \|_{L^{2}(\R^{3})}
        \; \leq  \; 
        C \; \| g \|_{H^{1}(\R^{3})} \quad \forall g \in H^{1}(\R^{3}) \; .
\end{equation*}
Using the eigenfunctions representation (\ref{g_eigenrepresentation}) of $\hat{\varrho}$'s integral kernel
we have
\begin{eqnarray}\label{eigen_summe_int_D1}
4 \pi \hat{D}^{k}_{\MM 1} \; g (x)
        & = &
        \sum\limits_{j \in \N} \; \lambda_{j} \; \partial_{x_{k}} \; \varphi_{j}(x) \; 
        \int\limits_{\R^{3}} \frac{\overline{\varphi_{j}(y)}}{|x-y|} \; g(y) \; dy \; .
\end{eqnarray}
Since $\varphi_{j} \in H^{1}(\R^{3})$ for $\hat{\varrho} \in Z$ (see (\ref{varphi_in_H1})) it remains to prove that
the second (integral) factor of
(\ref{eigen_summe_int_D1}) is in $L^\infty(\R^3_x)$.
Since we proceed here like
in the proof of Lemma {\ref{VHF_bounded}}, we only give the key
estimates. By a Sobolev embedding we have
\begin{eqnarray*}
\varphi_{j}, \, g
        & \in &
        H^{1}(\R^{3}) \hookrightarrow L^{p}(\R^{3}) \; , \quad 2 \leq  p \leq  6 \; .
\end{eqnarray*}
Hence, (\ref{use_gen_young}) directly gives
\begin{eqnarray*}
\big( \overline{\varphi_{j}} \cdot g \big) \ast \frac{1}{|x|}
        & \in &
        L^{q}(\R^{3}) \; , \quad 3 < q < \infty \; .
\end{eqnarray*}
Next we consider its spatial derivatives:
\begin{eqnarray*}
        \Big(\partial_{x_{k}} \; \overline{\varphi_{j}} \cdot g
        \; + \;
        \overline{\varphi_{j}} \cdot \partial_{x_{k}} \; g \Bigr)\ast \frac{1}{|x|} 
        & \in &
        L^{p}(\R^{3}) \; , \quad 3 < p < \infty \; 
\end{eqnarray*}
by proceeding as before.
In total we have 
$\; \big( \overline{\varphi_{j}} \cdot g \big) \ast 1/|x| 
         \in 
        W^{1,p}(\R^{3}) \; , \; 3 < p < \infty \; $, and the following estimate holds due to a Sobolev embedding:
\begin{eqnarray}{\label{convol_Linfty}}
\Big\| \big( \overline{\varphi_{j}} \cdot g \big) \ast \frac{1}{|x|} \Big\|_{L^{\infty}(\R^{3})} 
        & \leq &
        C \; 
        \big\| \varphi_{j} \big\|_{H^{1}(\R^{3})}
        \;  
        \big\| g \big\|_{H^{1}(\R^{3})} \; .
\end{eqnarray}
Now we can estimate (\ref{eigen_summe_int_D1}):
\begin{eqnarray*}
\big\| \hat{D}^{k}_{\MM 1} \; g \big\|_{L^{2}(\R^{3})}
        & \leq  &
        C \; 
        \sum\limits_{j \in \N} \; | \lambda_{j} | \; 
        \| \varphi_{j} \|^{2}_{H^{1}(\R^{3})} \; 
        \| g \|_{H^{1}(\R^{3})} \\
        & \leq &
        C \; \|\hat{\varrho}\|_{Z} \; \| g \|_{H^{1}(\R^{3})} \; .
\end{eqnarray*}
This proves
$\hat{D}^{k}_{\MM 1} \; \big( \hat{H}_{\MM 0}^{1/2} + I \big)^{-1} \in {\cal B}(L^{2}(\R^{3}))$, and hence
$\hat{D}^{k}_{\MM 1} \; \hat{\varrho} \; \partial_{x_j} \in {\cal I}_{\MM 1}$.\\
\\
The proof of (\ref{D2_in_I1}) is analogous to the proof of (\ref{D1_in_I1}):
In order to show
$\hat{D}^{k}_{\MM 2} \; \big( \hat{H}_{\MM 0}^{1/2} + I \big)^{-1} \in {\cal B}(L^{2}(\R^{3}))$, we consider
\begin{eqnarray}\label{eigen_summe_int_D2}
4 \pi \hat{D}^{k}_{\MM 2} \; g (x)
         = 
        \sum\limits_{j \in \N} \; \lambda_{j} \; \varphi_{j}(x) \; 
        \int\limits_{\R^{3}} \frac{\overline{\varphi_{j}(y)}}{|x-y|^{3}} \; (x_{k} - y_{k})
        \; g(y) \; dy, \;\; \forall \; g \in H^{1}(\R^{3}) .\;\;\;
\end{eqnarray}
To estimate the last integral we use $x_k /|x|^{3}  \in L^{3/2}_{w}(\R^{3})$
and
$\overline{\varphi_{j}} \cdot g \in L^{p}(\R^{3}) \; , \; 1 \leq  p \leq  3$, and get 
%
by the generalized Young inequality
\begin{eqnarray} \label{gen_young}
\Big\|  (\overline{\varphi_{j}} \cdot g ) * \frac{x_k}{|x|^{3}} \Big\|_{L^{p}(\R^{3})} 
        & \leq &
        C_p \; 
        \big\| \varphi_{j} \big\|_{H^{1}(\R^{3})}
        \;  
        \big\| g \big\|_{H^{1}(\R^{3})} \;,\; \frac{3}{2} < p < \infty .\;\;
\end{eqnarray}
Finally, 
\begin{eqnarray*}
4 \pi \big\| \hat{D}^{k}_{\MM 2} \; g \big\|_{L^{2}(\R^{3})}
        & \leq  &
        \sum\limits_{j \in \N} \; | \lambda_{j} | \; 
        \| \varphi_{j} \|_{L^{3}(\R^{3})} \; 
        \Big\|(\overline{\varphi_{j}} \cdot g ) * \frac{x_k}{|x|^{3}} \Big\|_{L^{6}(\R^{3})}  \\
        & \leq  &
        C \; \|\hat{\varrho}\|_{Z} \; \| g \|_{H^{1}(\R^{3})} \; ,
\end{eqnarray*}
which completes the proof of Part (a).\\
\\
$Part$ (b): By the same technique as before we shall prove $\partial^2_{x_{k}} \hat{V}^{\MM HF} \circ \hat{\varrho} \in {\cal
I}_1, \; k=1,2,3$.
We consider the kernel of the integral operator $\partial^2_{x_{k}}  \hat{V}^{\MM HF} $:
\begin{equation}\label{d2Vhf_ker}
\partial^2_{x_{k}} V^{\MM HF}(x,y) 
 =  \frac{1}{4 \pi}  \frac{\partial^2_{x_{k}} \varrho(x,y)}{ | x-y |} 
  - \frac{1}{2 \pi}  \partial_{x_{k}} \varrho (x,y) \frac{(x_k - y_k)}{ | x-y |^3}  
  - \frac{1}{4 \pi}  \varrho (x,y) \partial_{x_{k}} \left( \frac{x_k - y_k}{ | x-y |^3} \right),
\end{equation}
where the right hand side defines the kernels of three integral operators $\hat{A}^k_1$, $\hat{A}^k_2$,
$\hat{A}^k_3$ (in this
order).
To prove $\hat{A}^k_l \hat{\varrho} \in {\cal I}_1, \; k,l=1,2,3$, we shall use the factorization
\begin{equation*}
\hat{A}^k_l \hat{\varrho} = \hat{A}^k_l (\hat{H}_{\MM 0} + I)^{-1}(\hat{H}_{\MM 0}\hat{\varrho} + \hat{\varrho}) ,
\end{equation*}
with $\hat{H}_{\MM 0}\hat{\varrho} + \hat{\varrho} \in {\cal I}_1$ since $\hat{\varrho} \in Y$.
Hence, we have to show that $\hat{A}^k_l (\hat{H}_{\MM 0} + I)^{-1} \in {\cal B}(L^2(\R^3))$, that is
\begin{equation*}
\|\hat{A}^k_l g\|_{L^2(\R^3)} \le C \|g\|_{H^2(\R^3)} \;\; \forall f \in L^2(\R^3),
\end{equation*}
where $g \in H^2(\R^3)$ is defined as $g:=(\hat{H}_{\MM 0} + I)^{-1}f$ \;.\\
\\
{\it Case 1: Boundedness of} $\hat{A}^k_1 (\hat{H}_{\MM 0} + I)^{-1}$. In order to deal with the second
derivatives of $\varrho(x,y)$ in $\hat{A}^k_1$, we introduce two self-adjoint operators: 
\begin{eqnarray*}
\hat{\delta}_1:= \frac{1}{2}(\hat{H}_{\MM 0} \hat{\varrho} + \hat{\varrho}\hat{H}_{\MM 0} ) \in {\cal I}_1, \quad
\hat{\delta}_2:=\frac{1}{2i}(\hat{H}_{\MM 0} \hat{\varrho} - \hat{\varrho}\hat{H}_{\MM 0} ) \in {\cal I}_1,
\end{eqnarray*}
with their kernels satisfying $\delta_1, \delta_2\in
L^2(\R^6)$.
The eigenvector decomposition of $\hat{\delta}_1$ and $\hat{\delta}_1$ yields
\begin{eqnarray}{\label{delta12_def}}
\delta_1(x,y)  = \sum_{j \in \N} \mu_j \psi_j(x) \overline{\psi_j(y)} \quad \mbox{and} \quad
\delta_2(x,y)  = \sum_{j \in \N} \sigma_j \chi_j(x) \overline{\chi_j(y)} ,
\end{eqnarray}
with
\begin{eqnarray}
\|  \{\mu_j\}  \|_{l^1} &=& \frac{1}{2} \3norm \hat{H}_{\MM 0} \hat{\varrho} + \hat{\varrho}\hat{H}_{\MM 0} \3norm
_1 \le \3norm \hat{H}_{\MM 0} \hat{\varrho} \3norm_1 \;, {\label{lam_esti}}\\
\|  \{\sigma_j\}  \|_{l^1} &=& \frac{1}{2} \3norm \hat{H}_{\MM 0} \hat{\varrho} - \hat{\varrho}\hat{H}_{\MM 0}
\3norm_1 \le \3norm \hat{H}_{\MM 0} \hat{\varrho} \3norm_1 \; , {\label{mu_esti}}
\end{eqnarray}
and $\{\psi_j\},\;\{\chi_j\}$ are orthonormal systems in $L^2(\R^3)$. For each
 $g \in H^2(\R^3)\hookrightarrow L^{q}(\R^3)$, $2 \le q \le
\infty$, we consider
\begin{eqnarray*}
f_j &:=& (\overline{\psi_j}g) * \frac{1}{|x|} \in L^{p}(\R^3), \;\; 3<p<\infty  \;\; \mbox{and} \\
\nabla f_j &:=& -(\overline{\psi_j}g) * \frac{x_j}{|x|^3}  \in L^{q}(\R^3), \;\;\frac{3}{2}<q \le 6 \;.
\end{eqnarray*}
A Sobolev embedding then implies
\begin{eqnarray}{\label{fj_estim}}
\|f_j \|_{L^{\infty}(\R^3)} \le  C \|\psi_j\|_{L^2(\R^3)} \|g\|_{H^2(\R^3)} \quad .
\end{eqnarray}
The same argument holds for $h_j:=(\overline{\chi_j}g) * 1/|x| :$
\begin{eqnarray}{\label{hj_estim}}
\|h_j \|_{L^{\infty}(\R^3)} \le  C \|\chi_j\|_{L^2(\R^3)} \; \|g\|_{H^2(\R^3)} \quad .
\end{eqnarray}
Using $\hat{H}_{\MM 0} \hat{\varrho} = \hat{\delta}_1 + i \hat{\delta}_2$ and ({\ref{delta12_def}}) - ({\ref{hj_estim}}) we estimate:\\[1ex]
$
2 \pi \| \sum_{k=1}^3 \hat{A}^k_1 g\|_{L^2(\R^3)} 
\quad =\quad  \displaystyle\frac{1}{2} \Big{\|}  \int_{\R^3} \frac{\Delta_x \varrho (x,y)}{|x-y|}g(y)dy \Big{\|}_{L^2(\R^3)}\\ [1ex]
=\quad  \Big{\|} \displaystyle\sum_{j \in \N}   \mu_j  \psi_j (x) \int \frac{\overline{\psi_j (y)}g(y)}{|x-y|}dy 
+  i \sum_{j \in \N}  \sigma_j \chi_j (x) \int \frac{\overline{\chi_j (y)}g(y)}{|x-y|}dy \Big{\|}_{L^2(\R^3)}\\ [1ex]
\le\quad  \displaystyle\sum_{j \in \N} | \mu_j| \;
 \| \psi_j \|_{L^2(\R^3)} \|f_j\|_{L^\infty(\R^3)} 
+ \sum_{j \in \N} | \sigma_j| \;
 \| \chi_j \|_{L^2(\R^3)} \|h_j\|_{L^\infty(\R^3)} \\[1ex]
\le\;  C \displaystyle\sum_{j \in \N} (| \mu_j | + | \sigma_j |) \|g\|_{H^2(\R^3)} 
\;\le\; C  \3norm \hat{H}_{\MM 0} \hat{\varrho}\3norm_1 \|g\|_{H^2(\R^3)} 
\;\le\;  C  \|\hat{\varrho}\|_{ Y} \|g\|_{H^2(\R^3)} \;.
$\\[1ex]
\\
{\it Case 2: Boundedness of} $\hat{A}^k_2 (\hat{H}_{\MM 0} + I)^{-1}$. 
With the eigenfunction decomposition of $\hat{\varrho}$ we have
\begin{equation*}
\hat{A}^k_2 g = -\frac{1}{2 \pi} \sum_{j \in \N} \lambda_j \; \partial_{x_k} \varphi_j(x) \int_{\R^3}
\overline{\varphi_j(y)} \; \frac{x_k - y_k}{ | x-y |^3} \; g(y) \; dy \; .
\end{equation*}
As in ({\ref{gen_young}}) we have $(\overline{\varphi_j}g) * (x_k / | x|^3) \in L^p(\R^3) ,\;
3/2 < p <\infty $, 
and its spatial derivatives are in $L^q(\R^3), \; 3/2 <q \le 6 $. Hence we have
\begin{eqnarray*}
\Big\|(\overline{\varphi_j}g) * \frac{x_k}{ | x|^3} \Big\|_{L^\infty(\R^3)}
 \le  C \|\varphi_j\|_{H^1(\R^3)} \|g\|_{H^2(\R^3)} \; .
\end{eqnarray*}
So we obtain the desired result
\begin{eqnarray*}
\| \hat{A}^i_2 g\|_{L^2(\R^3)} 
 \le  C \sum_{j \in \N} | \lambda_j | \; \|\varphi_j\|^2_{H^1(\R^3)} \|g\|_{H^2(\R^3)} 
\; \le  C  \|\hat{\varrho}\|_{Y} \|g\|_{H^2(\R^3)} \; .
\end{eqnarray*}
\\
{\it Case 3: Boundedness of} $\hat{A}^k_3 (\hat{H}_{\MM 0} + I)^{-1}$. We rewrite the last term of ({\ref{d2Vhf_ker}}) as
\begin{eqnarray*}
\hat{A}^k_3 g(x) 
& =& -\frac{1}{4 \pi} \sum_{j \in \N} \lambda_j  \varphi_j(x)
\Big( ( \overline{\varphi_j}g )* \partial_{x_{k}} \big( \frac{x_k}{ | x |^3} \big) \Big) 
\\& = &
-\frac{1}{4 \pi} \sum_{j \in \N} \lambda_j  \varphi_j(x)
\Big( \partial_{x_{k}}( \overline{\varphi_j}g )*  \frac{x_k}{ | x |^3}  \Big)
\end{eqnarray*}
and estimate as in Case 2:
\begin{eqnarray*} 
| \partial_{x_{k}}(\overline{\varphi_j} g)| * \frac{x_k}{ | x|^3} \in L^p(\R^3) \quad
\frac{3}{2} < p \le 6 \; ,
\end{eqnarray*}
and hence
\begin{eqnarray*}
\| \hat{A}^k_3 g\|_{L^2(\R^3)} & \le & C \sum_{j \in \N} | \lambda_j | \;
\| \varphi_j \|_{L^6(\R^3)} \Big\| \partial_{x_{k}}(\overline{\varphi_j} g) * \frac{x_k}{ |
x|^3} \Big\|_{L^3(\R^3)} \\
& \le & C \sum_{j \in \N} | \lambda_j | \; \|\varphi_j\|^2_{H^1(\R^3)} \; \|g\|_{H^2(\R^3)} 
\quad \le \; C  \|\hat{\varrho}\|_{ Y} \|g\|_{H^2(\R^3)} \; .
\end{eqnarray*}
Summarising we have proved $\hat{H}_{\MM 0}\hat{V}^{\MM HF}(\hat{H}_{\MM 0}+I)^{-1} \in {\cal B}(L^2(\R^3))$ and the assertion (b) follows.
\qed
\vspace{3mm}


\begin{lemma}\label{F_herm}
$F(\hat{\varrho})$ is Hermitian 
for a given $\hat{\varrho} \in Z$. 
\end{lemma}

\proof Let $\hat{\varrho} \in Z$, hence it is Hermitian. 
The assertion is a simple consequence of
\begin{eqnarray*}
        V^{\MM H}[\hat{\varrho}](x) \in \R, \quad  
        V^{\MM HF}[\hat{\varrho}](x,z)= \overline{V^{\MM
        HF}[\hat{\varrho}](z,x)}, 
\end{eqnarray*}
i.e.\ the self-adjointness of $\hat{V}^{\MM H}[\hat{\varrho}]$ and $\hat{V}^{\MM HF}[\hat{\varrho}]$ (cf.\ (\ref{def_h0_F})).
\qed
\vspace{3mm}

The analogous
properties of the previous Lemmata for $\hat{V}^{\MM H}[\hat{\varrho}]$ can be proved with the following generalization of the Lieb-Thirring inequalities
(cf.\ Refs.\ \cite{LiPa} and \cite{Arn96}, Theorem A.3.):
\begin{eqnarray*}
\|n\|_{L^q(\R^3)} &\le& C_{\MM q} \3norm \hat{\varrho} \3norm_1^{\alpha} 
\;(\mbox{Tr}(\hat{H}_{\MM 0}^{1/2} |\hat{\varrho}| \hat{H}_{\MM 0}^{1/2}))^{1-\alpha} , \; 1 \le q \le 3,\\
\|\nabla n\|_{L^r(\R^3)} &\le& C_{\MM r} \3norm \hat{\varrho} \3norm_1^{\beta} 
\;(\mbox{Tr}(\hat{H}_{\MM 0}^{1/2} |\hat{\varrho}| \hat{H}_{\MM 0}^{1/2}))^{1-\beta} , \; 1 \le r \le 3/2, \\
&&\alpha = (3-q)/2q ,\quad \beta = (3-2r)/2r,
\end{eqnarray*}
As a consequence of the Lemmata \ref{VHF_bounded}\; - \;\ref{F_herm} we then obtain:

\begin{corollary}\label{theore_F_lipsch}
$F$ maps $Z$ (resp. $Y$) into itself and is locally Lipschitz in $Z$ (resp. $Y$).
\end{corollary}
\vspace{3mm}
Using standard perturbation results (Ref.\ \cite{Paz83}, Theorem 6.4.1,) the local Lipschitz continuity of the function 
$F(\hat{\varrho})$ guarantees the existence of a local-in-time
solution of (\ref{HF_system}).
\begin{theorem}\label{th_loc_sol}
Let $\hat{\varrho}^{\MM I} \in Z$. \\
\emph{(a)} Then the Hartree-Fock system (\ref{HF_system}) has a unique mild solution $\hat{\varrho}
\in$ \linebreak $C([0,t_{max});Z)$ with a
potential $\hat{V}^{\MM H} - \hat{V}^{\MM HF} \in C([0,t_{max});{\cal B}(L^2(\R^3)))$. Moreover, if the maximum time interval is finite, i.e.\ $t_{max} <
\infty$, then \\ 
\;$\lim_{t \uparrow t_{max}} \|\hat{\varrho}\|_{Z} = \infty$. \\
\emph{(b)} In case $\hat{\varrho}^{\MM I} \in Y \subset {\cal D}(h_{\MM 0})$,  $\hat{\varrho}$ is a classical solution with  $\hat{\varrho}
\in C([0,t_{max});Y) \cap C^1([0,t_{max});{\cal I}_1)$.\\
\emph{(c)} For all $0<t_1 < t_{max}$ the map $ \hat{\varrho}^{\MM I} \mapsto \hat{\varrho}(t)$ is
Lipschitz continuous on some (small enough) ball $\{\|\hat{\varrho}-\hat{\varrho}_{\MM 0}\|_{Z} <
\varepsilon(t_1)\}\subset Z$, 
uniformly for $0 \le t \le t_1$.
\end{theorem}

\begin{remark}
Since equation (\ref{HF_system}) is in Hamiltonian form, it preserves positivity:
if the initial value $\hat{\varrho}^{\MM I} \ge 0$, then $\hat{\varrho}(t) \ge 0 \quad \forall t \in [0,
t_{max})$.\\
\\
From now on we will restrict our analysis to positive density matrix operators, which represent physical quantum states.
\end{remark}


\renewcommand{\thesection}{\Roman{section}}
\section{Global-In-Time Solution}\label{sec:global_time_solution}
\renewcommand{\thesection}{\arabic{section}}

%
%

In order to prove the global-in-time existence of solutions to equation 
(\ref{HF_system}), we shall derive an $a$ $priori$ estimate for the kinetic energy (cf.\ Lemma \ref{lemma_boun_kin_en}).
This estimate is a consequence of the conservative character of the problem.
More precisely, we show that the total charge and the total energy
are conserved by the local-in-time solution $\hat{\varrho}$ of the Hartree-Fock system.
On a formal level this is well known since Dirac \cite{Di30}, but we shall need here a rigorous proof.
We recall that the total charge corresponds, by definition, to the quantity $\3norm \hat{\varrho} \3norm_1$. And the 
total energy of the Hartree-Fock system is given by 
\begin{eqnarray*}
E_{tot}(\hat{\varrho}) := E_{kin}(\hat{\varrho}) + E_{pot}(\hat{\varrho}) \; ,
\end{eqnarray*}
where the kinetic energy $E_{kin}(\hat{\varrho})$ is defined in (\ref{def_E_kin}) and the potential energy equals
\begin{eqnarray*}
E_{pot}(\hat{\varrho})  &:=& \frac{1}{2} \mbox{Tr} ( \hat{\varrho} \hat{V}^{\MM H}) -\frac{1}{2} 
\mbox{Tr}(\hat{\varrho} \hat{V}^{\MM HF})\;.
\end{eqnarray*}
\vspace{3mm}

%
\begin{lemma}\label{lemma_cons_en}
Let $\hat{\varrho}^{\MM I} \in Z$ and $\hat{\varrho}^{\MM I} \ge 0$ in (\ref{HF_system}). Then the local solution $\hat{\varrho}$
 of Theorem \ref{th_loc_sol} satisfies
\begin{eqnarray*}
\3norm \hat{\varrho}(t)\3norm_1 = \3norm \hat{\varrho}^{\MM I} \3norm_1 ,\quad E_{tot}(\hat{\varrho}(t)) = E_{tot}(\hat{\varrho}^{\MM I})
\quad \forall t \in [0,t_{max}) \;.
\end{eqnarray*}
\end{lemma}

\proof
The conservation of the total charge for (\ref{HF_system}) can be proved by writing  the integral equation associated to (\ref{HF_system})
in terms of the semigroup $\{G_{\MM 0}(s)\}$
and the perturbation $F(\hat{\varrho})$:
\begin{eqnarray}\label{explic_sol}
\hat{\varrho}(t) =  G_{\MM 0} (t)\hat{\varrho}^{\MM I} + \int_0^t G_{\MM 0} (t-s)\; F(\hat{\varrho}(s)) \;ds  , \quad 0\le t < t_{max}
\end{eqnarray}
and taking the trace. 
For $\hat{\varrho}(s) \in Z$ we have $\hat{V}^{\MM H}(s) - \hat{V}^{\MM HF}(s)  \in {\cal B}(L^2(\R^3))$. Using the representation (\ref{rep_isom_evo_group}) 
of $G_{\MM 0} (t)$ and the properties of the trace
we obtain Tr$F(\hat{\varrho}(s))=0$, which implies 
 Tr$\hat{\varrho}(t) =  \mbox{Tr}\hat{\varrho}^{\MM I} \;  \forall t \in [0,t_{max})$.
%

Concerning the conservation of the total energy we first restrict the analysis to $\hat{\varrho}^{\MM I} \in Y$, 
which by Theorem \ref{th_loc_sol} implies the local-in-time existence of a classical solution $\hat{\varrho}(t)$.
The general case $\hat{\varrho}^{\MM I} \in Z$ will be then derived by a density argument.
For $\hat{\varrho} \in Y$ all of the following manipulations are well defined. Hence we apply the operator $\hat{H}_{\MM 0}$ to both sides of (\ref{explic_sol}) and then take traces.
Commuting $G_{\MM 0}(t)$ with $\hat{H}_{\MM 0}$, using $\hat{V}^{\MM H}(s)- \hat{V}^{\MM HF}(s)  \in {\cal B}(L^2(\R^3))$,
and using the cyclicity of the trace, we obtain (as an extension of Lemma 3.7 in Ref.\ \cite{Arn96}):
\begin{equation}
\begin{array}{rl}
\mbox{Tr}(\hat{H}_{\MM 0} \hat{\varrho}(t)) =&  
 \mbox{Tr}(\hat{H}_{\MM 0} \hat{\varrho}^{\MM I}) 
-i \displaystyle\int_0^t  \mbox{Tr} \big(\hat{H}_{\MM 0} \hat{V}^{\MM H} [\hat{\varrho}]\;\hat{\varrho}(s) - \hat{H}_{\MM 0} \hat{\varrho}\hat{V}^{\MM H} [\hat{\varrho}](s)\big) \;ds
\\
&+i \displaystyle\int_0^t  \mbox{Tr} \big(\hat{H}_{\MM 0} \hat{V}^{\MM HF} [\hat{\varrho}]\;\hat{\varrho}(s) - \hat{H}_{\MM 0} \hat{\varrho}\hat{V}^{\MM
HF} [\hat{\varrho}](s)\big) \;ds \; 
\\
=& \mbox{Tr}(\hat{H}_{\MM 0} \hat{\varrho}^{\MM I}) 
+i \displaystyle\int_0^t  \mbox{Tr} \big(\hat{V}^{\MM H} [\hat{\varrho}]\;[\hat{H}_{\MM 0}, \hat{\varrho}(s) ]\big) \;ds  
\\&-i \displaystyle\int_0^t  \mbox{Tr} \big(\hat{V}^{\MM HF}[\hat{\varrho}]\;[\hat{H}_{\MM 0}, \hat{\varrho}(s) ]\big) \;ds 
\\
=& \mbox{Tr}(\hat{H}_{\MM 0} \hat{\varrho}^{\MM I}) 
- \displaystyle\int_0^t  \mbox{Tr} \big(\hat{V}^{\MM H} [\hat{\varrho}]\;\hat{\varrho}_t(s)\big) \;ds  
+ \displaystyle\int_0^t  \mbox{Tr} \big(\hat{V}^{\MM HF}[\hat{\varrho}]\;\hat{\varrho}_t(s)\big) \;ds  ,  \label{tr_eq}
\end{array}
\end{equation}
where we used (\ref{HF_system}) in the last step.
Applying an integration by parts in $t$ (possible since $\hat{\varrho}_t \in C([0,t_{max}); {\cal I}_{\MM 1}$),
the second term on the right hand side\ of (\ref{tr_eq}) becomes
\begin{equation}
\begin{array}{l}
\displaystyle\int_0^t  \mbox{Tr} \big(\hat{V}^{\MM H} [\hat{\varrho}]\;\hat{\varrho}_t(s)\big) \;ds 
\\ = - \mbox{Tr} \big( V^{\MM H} [\hat{\varrho}](0) \; \hat{\varrho}^{\MM I}\big) 
+ \mbox{Tr}\big( V^{\MM H} [\hat{\varrho}](t) \; \hat{\varrho}(t) \big) 
- \displaystyle\int_0^t  \mbox{Tr} \big(V^{\MM H}_t [\hat{\varrho}](s)\;\hat{\varrho}(s)\big) \;ds . \label{tr_eq_VH}
\end{array}
\end{equation}
For this last integral term we note that $n_t \in C([0,t_{max});L^1(\R^3))$ implies, by the
Hardy-Littlewood-Sobolev inequality, that $V^{\MM H}_t[\hat{\varrho}] = (1/4 \pi |x|)*n_t
\in C([0,t_{max});$ $L^{3}_w(\R^3))$.  Using $\hat{\varrho} \in C([0,t_{max});Y)$ and techniques like
in the proof of Lemma \ref{lemma_in_I} one easily obtains $\hat{V}^{\MM H}_t[\hat{\varrho}]
\hat{\varrho} \in C([0,t_{max});{\cal I}_{\MM 1})$ and the following equivalence holds by using the
Poisson equation (\ref{poisson_eq}) and two integration by parts (in $x$):
\begin{eqnarray}\label{tr_eq_VH2}
\mbox{Tr} (\hat{V}^{\MM H}_t[\hat{\varrho}] \; \hat{\varrho}(s)) 
&=& - \int_{\R^3} V^{\MM H}_t[\hat{\varrho}](x,s) \; \Delta V^{\MM H}[\hat{\varrho}](x,s) \; dx
\nonumber\\
&=& \frac{1}{2}\frac{d}{dt}\int_{\R^3} | \nabla V^{\MM H}[\hat{\varrho}](x,s)|^2 \; dx \nonumber
\\&=& \frac{1}{2}\frac{d}{dt}\int_{\R^3} V^{\MM H}[\hat{\varrho}](x,s) \; n(x,s) \; dx \nonumber\\
&=& \frac{1}{2} \frac{d}{dt} \mbox{Tr} (\hat{V}^{\MM H}[\hat{\varrho}]\hat{\varrho}(s)) \; . \quad
\end{eqnarray}
Since $\hat{\varrho}_t \in C([0,t_{max});{\cal I}_{\MM 1})$ and $\hat{V}^{\MM HF}[\hat{\varrho}] \in {\cal B}(L^2(\R^3))$
(cf.\ Lemma
\ref{VHF_bounded}),
the last term of (\ref{tr_eq}) is well defined.
Using (\ref{kernel_pres_HF}) we finally compute it:
\begin{eqnarray}\label{tr_eq_VHF}
\mbox{Tr} \big(\hat{V}^{\MM HF} [\hat{\varrho}]\;\hat{\varrho}_t(s)\big)
&=&  \frac{1}{4 \pi}\Big( \int_{\R^3} dx \int_{\R^3} dz \; \frac{\varrho(x,z,s)}{|x-z|}\; \varrho_t(z,x,s) \Big) \nonumber \\
&=&  \frac{1}{4 \pi} \; \frac{1}{2}\frac{d}{dt} \Big( \int_{\R^3} dx \int_{\R^3} dz \; \frac{\varrho(x,z,s)\; \varrho(z,x,s)}{|x-z|} \Big) \nonumber \\
&=&  \frac{1}{2} \frac{d}{dt} \mbox{Tr} \big(\hat{V}^{\MM HF} [\hat{\varrho}]\;\hat{\varrho}(s)\big) \;.
\end{eqnarray}
Plugging the expressions (\ref{tr_eq_VH}) - (\ref{tr_eq_VHF}) into (\ref{tr_eq}) we obtain 
\begin{eqnarray*}
\mbox{Tr}(\hat{H}_{\MM 0} \hat{\varrho}(t)) 
+ \frac{1}{2} \mbox{Tr}\big( \hat{V}^{\MM H} [\hat{\varrho}] \; \hat{\varrho}(t) \big) 
- \frac{1}{2} \mbox{Tr}\big( \hat{V}^{\MM HF} [\hat{\varrho}] \; \hat{\varrho}(t) \big) \quad\quad
\\[1ex]  = \; \mbox{Tr}(\hat{H}_{\MM 0} \hat{\varrho}^{\MM I}) +
\frac{1}{2} \mbox{Tr} \big( \hat{V}^{\MM H} [\hat{\varrho}](0) \; \hat{\varrho}^{\MM I}\big)
- \frac{1}{2} \mbox{Tr} \big( \hat{V}^{\MM HF} [\hat{\varrho}](0) \; \hat{\varrho}^{\MM I}\big) \; ,
\end{eqnarray*}
which proves the assertion for $\hat{\varrho}^{\MM I} \in Y$.\\

Next we consider an initial condition $\hat{\varrho}^{\MM I}\in Z$ with $\hat{\varrho}^{\MM I}\ge
0$. Due to Lemma \ref{Y_in_Z}(c) it can be approximated by a sequence $\{\hat{\varrho}_n^{\MM I}\} \subset
Y$ such that $\hat{\varrho}_n^{\MM I} \stackrel{Z}{\longrightarrow} \hat{\varrho}^{\MM I}$,  with
$\hat{\varrho}_n^{\MM I} \ge 0$, $\forall n \in \N$. For any $0<t_1<t_{max}$, the corresponding
trajectories $\hat{\varrho}_n(t),\, 0\le t\le t_1$ lie in $Y$ (cf.\ Theorem \ref{th_loc_sol}(b)) and they 
converge in $Z$ to $\hat{\varrho}(t)$, uniformly on $[0,t_1]$ (cf.\ Theorem \ref{th_loc_sol}(c)).
This implies convergence of the kinetic energy of $\hat{\varrho}_n=\hat{\varrho}_n(t)$ $\forall
t\in[0,t_1]$:
$$
  E_{kin}(\hat{\varrho}_n) = \3norm \hat{H}_{\MM 0}^{1/2} \hat{\varrho}_n \hat{H}_{\MM 0}^{1/2}
  \3norm_1 \stackrel{n \rightarrow \infty}{\longrightarrow}  \3norm \hat{H}_{\MM 0}^{1/2}
  \hat{\varrho} \hat{H}_{\MM 0}^{1/2} \3norm_1 = E_{kin}(\hat{\varrho}) \; . 
$$
To prove convergence of the Hartree-Fock potential energy we estimate:
\begin{eqnarray}\label{h5}
  \big|\mbox{Tr} ( \hat{V}^{\MM HF}\!\!\!\!\!\!&[&\!\!\!\!\!\!\hat{\varrho}_n] \circ \hat{\varrho}_n) 
  - \mbox{Tr} ( \hat{V}^{\MM HF}[\hat{\varrho}] \circ \hat{\varrho})\big| \nonumber \\  
  &\le &\| \hat{V}^{\MM HF}[\hat{\varrho}_n] - \hat{V}^{\MM HF}[\hat{\varrho}] \|_{{\cal B}(L^2(\R^3))} \; \3norm \hat{\varrho}_n \3norm_1 +
\| \hat{V}^{\MM HF}[\hat{\varrho}] \|_{{\cal B}(L^2(\R^3))} \3norm \hat{\varrho}_n - \hat{\varrho} \3norm_1 \nonumber \\
&\le& C \| \hat{\varrho} \|_{ Z} \; \| \hat{\varrho}_n - \hat{\varrho} \|_{ Z} \stackrel{n \rightarrow \infty}{\longrightarrow} \; 0, 
\end{eqnarray}
where we used the estimate $\| \hat{V}^{\MM HF}[\hat{\varrho}_n - \hat{\varrho}] \|_{{\cal B}(L^2(\R^3))} \le C \| \hat{\varrho}_n - \hat{\varrho}
\|_{ Z}$ obtained in the proof of Lemma \ref{VHF_bounded}.\\
The convergence of the Hartree potential energy, i.e.\ 
$$
  \mbox{Tr} ( \hat{V}^{\MM H}[\hat{\varrho}_n] \circ \hat{\varrho}_n)
  \stackrel{n \rightarrow \infty}{\longrightarrow}
  \mbox{Tr} ( \hat{V}^{\MM H}[\hat{\varrho}] \circ \hat{\varrho})
$$
is obtained analogously by using the estimate $\| V^{\MM H}[\hat{\varrho}]\|_{L^{\infty}(\R^3)}  \le C 
\| \hat{\varrho} \|_{ Z}$ derived in Equation (3.46) of Ref.\ \cite{Arn96}.\\ 
The assertion of the lemma now follows from 
\begin{eqnarray*}
  E_{kin}(\hat{\varrho}_n(t)) + E_{pot}(\hat{\varrho}_n(t)) 
  = E_{kin}(\hat{\varrho}^{\MM I}_n) + E_{pot}(\hat{\varrho}^{\MM I}_n), \quad
  \forall n \in \N,\;t\in [0,t_1]
\end{eqnarray*}
in the limit $n \rightarrow \infty$. \qed
\vspace{3mm}

\begin{lemma}\label{lemma_boun_kin_en}
Let $\hat{\varrho}^{\MM I} \in Z$ and $\hat{\varrho}^{\MM I} \ge 0$. Then the kinetic energy of the Hartree-Fock system (\ref{HF_system}) is bounded
$\forall t \in [0,t_{max})$:
\begin{eqnarray}{\label{boun_kin_en}}
0 \le E_{kin}(\hat{\varrho}(t)) \le E_{tot}(\hat{\varrho}^{\MM I}) \; .
\end{eqnarray}
\end{lemma}
\proof
The kinetic energy $E_{kin}(\hat{\varrho}(t))$ is non negative by definition and equals
$E_{tot}(\hat{\varrho}^{\MM I})-E_{pot}(\hat{\varrho}(t))$
by Lemma \ref{lemma_boun_kin_en}.
To complete the proof we have to show that $E_{pot}(\hat{\varrho}(t)) \ge 0$,  $\forall t \in [0,t_{max})$.
\begin{eqnarray*}
2 E_{pot}(\hat{\varrho}(t)) &=& \mbox{Tr} ( \hat{\varrho} \hat{V}^{\MM H}) - \mbox{Tr}(\hat{\varrho} \hat{V}^{\MM HF})\\
&=& \sum_{j \in \N} \lambda_j \int_{\R^3} dx \; |\psi_j(x)|^2 \; \Big( \frac{1}{4 \pi} \int_{\R^3} dz \; \frac{n(z)}{|x-z|} \Big) \\
&& - \sum_{j \in \N} \lambda_j \int_{\R^3} dx \; \overline{\psi_j(x)} \; \Big(\frac{1}{4 \pi} \int_{\R^3} dz \;
\frac{\varrho(x,z)}{|x-z|}\psi_j(z) \Big)\\
&=& \int_{\R^3}dx\int_{\R^3}dz \; \frac{n(x) n(z)}{4 \pi |x-z|} - \int_{\R^3}dx\int_{\R^3}dz \;
\frac{|\varrho(x,z)|^2}{4 \pi |x-z|} \; ,
\end{eqnarray*}
which is non-negative since, by the Schwarz inequality, we have:
\begin{eqnarray*}
|\varrho(x,z)|^2 &=& \Big |\sum_{j \in \N} \lambda_j^{1/2}\; \psi_j(x) \; \lambda_j^{1/2} \; \overline{\psi_j}(z) \Big|^2\\
&\le& \Big(\sum_{j \in \N} \lambda_j \; |\psi_j(x)|^2 \Big) \Big(\sum_{j \in \N} \lambda_j \; |\psi_j(z)|^2 \Big)\\
&=& n(x)\; n(z) \; .
\end{eqnarray*}
\qed
\vspace{3mm}

From Lemma \ref{lemma_cons_en} and \ref{lemma_boun_kin_en} we conclude that
the local-in-time solution of Theorem \ref{th_loc_sol} can be extended to
$t_{max}=\infty$ and the global solution theorem follows:

\begin{theorem}{\label{th_glob_sol}}
Let $\hat{\varrho}^{\MM I} \in Z$  and $\hat{\varrho}^{\MM I} \ge 0$. Then the Hartree-Fock system (\ref{HF_system})
has a unique global mild solution $\hat{\varrho}\in C([0,\infty);Z)$ with a
potential $\hat{V}^{\MM H} - \hat{V}^{\MM HF} \in C([0,\infty);{\cal B}(L^2(\R^3)))$. \\
In case $\hat{\varrho}^{\MM I} \in Y$ and $\hat{\varrho}^{\MM I} \ge 0$,  $\hat{\varrho}$ is a global classical solution 
with $\hat{\varrho}\in C([0,\infty);Y)\cap C^1([0,\infty);{\cal I}_1) $.
\end{theorem}
%
%
\vskip 1 cm
\textbf{Acknowledgments:}
\newpar
The first author was partially supported by the European Union research network
\emph{HYKE} and the DFG-project AR$277/3$-$2$. The second author was supported
by the DFG-Graduiertenkolleg: \emph{Nichtlineare kontinuierliche Systeme und
deren  Untersuchung mit numerischen, qualitativen und experimentellen
Methoden}.





\newpage
\begin{thebibliography}{BoDPFa74}

\bibitem[1]{Di30}
P. A. M.\ Dirac,
``Note on exchange phenomena in the Thomas atom,''
Proc.\ Cambridge Philos.\ Soc. {\bf 26}, 376--385 (1930).


\bibitem[2]{Sla50}
J. C.\ Slater,
``A simplification of the Hartree-Fock method,''
Phys.\ Rev.~{\bf 81}(3), 385--390 (1951).


\bibitem[3]{DaLi90}
R.\ Dautray and J. L.\ Lions, {\it Mathematical Analysis and Numerical
Methods for Science and Technology\/}, Vol. 1 (Springer-Verlag,
Berlin, New York, 1990).


\bibitem[4]{Thi80} 
W.\ Thirring, {\it Lehrbuch der mathematischen Physik,
Vol.~4: Quantenmechanik gro\ss er Systeme} (Springer-Verlag, Wien, 1980).


\bibitem[5]{Arn96}
A.\ Arnold, ``Self-Consistent Relaxation-Time Models In
Quantum Mechanics,'' Comm. Part. Diff. Eq. {\bf 21}, No. 3\&4, 473--506 (1996).


\bibitem[6]{LiPa}
P. L.\ Lions and T.\ Paul, 
``Sur les mesures de Wigner,''
Rev. Math. Iberoam. {\bf 9}, 553--618 (1993).


\bibitem[7]{Mar89}
P. A.\ Markowich,
``On the equivalence of the Schr\"odinger equation and the quantum
Liouville equation,''
Math. Methods Appl. Sci. {\bf 11}, 459--469 (1989).


\bibitem[8]{ChGl75}
J. M.\ Chadam and R. T.\ Glassey,
``Global existence of solutions to the Cauchy problem for time-dependent Hartree
equations,''
J. Math. Phys. {\bf 16}, 1122--1130 (1975).


\bibitem[9]{Gas99}
I.\ Gasser,
``On Hartree-Fock Systems,''
VSLI Design {\bf 9}, 357--364 (1999).


\bibitem[10]{Cast97}
F.\ Castella,
``${L}^2$ solutions to the Schr\"odinger-Poisson system:
Existence, uniqueness, time behaviour, and smoothing effect,''
Math.\ Models Meth.\ Appl.\ Sci. {\bf 7}, 1051--1083 (1997).


\bibitem[11]{GIMS98}
I. Gasser, R. Illner, P. A. Markowich, and C. Schmeiser,
``Semiclassical, $t\to\infty$ asymptotics and dispersive effects for Hartree-Fock
systems,''
Math. Mod. Num. Anal. {\bf 32}, 699--713 (1998).


\bibitem [12] {DIL} J. Dolbeault, R. Illner, and H. Lange,
``On asymmetric quasiperiodic solutions of Hartree-Fock systems,''
J. Differential Equations {\bf 178}, No. 2, 314--324 (2002).


\bibitem[13]{BoDPFa74}
A.\ Bove, G.\ DaPrato, and G.\ Fano,
``An Existence Proof for the Hartree-Fock Time-dependent Problem with Bounded
Two-Body Interaction,''
Commun. Math. Phys. {\bf 37}, 183--191 (1974).


\bibitem [14] {AlFa} 
R. Alicki and M. Fannes, {\it Quantum dynamical systems}  
(Oxford University Press, 2001).


\bibitem[15]{Da} 
E. B.\ Davies, {\it Quantum Theory of Open Systems}
(Academic Press, 1976).


\bibitem[16]{ArSp03}
A.\ Arnold and C.\ Sparber,
``Conservative Quantum Dynamical Semigroups for a Class
of Mean Field Master Equations,'' (preprint, 2003).


\bibitem[17]{Arn97}
A.\ Arnold,
``The relaxation-time von Neumann-Poisson equation.
ZAMM,'' {\bf 76}(S2), 293--296 (1996).
Proceedings of ICIAM 95, Hamburg (1995), Oskar Mahrenholtz, Reinhard
Mennicken (eds.).


\bibitem [18] {Li} G.\ Lindblad, 
``On the generators of quantum mechanical semigroups,''
Comm. Math. Phys. {\bf 48}, 119--130 (1976).


\bibitem[19]{RSI}
M.\ Reed and B.\ Simon, {\it Methods of Modern Mathematical Physics, 
Vol.~I: Functional Analysis,\/} (Academic Press, New York and London, 1972).


\bibitem[20]{Sim78}
B.\ Simon, {\it Trace Ideals and Applications,\/} (Cambridge University 
Press, 1978).


\bibitem[21]{Paz83}
A.\ Pazy, {\it Semigroups of linear operators and applications to partial 
differential equations} (Springer--Verlag, New York, 1983).


\bibitem[22]{DaLi92}
R.\ Dautray and J. L.\ Lions, {\it Mathematical Analysis and 
Numerical Methods for Science and Technology\/}, Vol. 5 (Springer-Verlag,
Berlin, New York, 1992).


\bibitem[23]{RSII}
M.\ Reed and B.\ Simon, {\it Methods of Modern Mathematical Physics, 
Vol.~II: Fourier analysis, self-adjointness,\/} (Academic Press, New York, San 
Francisco, and London, 1975).


\bibitem[24]{Ste70}
E. M.\ Stein, {\it Singular Integrals and Differentiability Properties of
Functions,\/} (Princeton University Press, 1970).

\end{thebibliography}
\end{document}